\documentclass{aa}
\usepackage{graphicx}
\usepackage{txfonts}
\usepackage{amssymb}
\usepackage{lscape}
\begin{document}

\def\mpc{h^{-1} {\rm{Mpc}}} 
\def\kpc{h^{-1} {\rm{kpc}}}
\newcommand{\mincir}{\raise
-2.truept\hbox{\rlap{\hbox{$\sim$}}\raise5.truept\hbox{$<$}\ }}
\newcommand{\magcir}{\raise
-2.truept\hbox{\rlap{\hbox{$\sim$}}\raise5.truept\hbox{$>$}\ }}

\title{The XXL Survey\thanks{Based on observations obtained with XMM-Newton, an ESA science 
mission with instruments and contributions directly funded by
ESA Member States and NASA. Based on observations obtained with the William Herschel telescope during semester 13B.}} 

\subtitle{XII. Optical spectroscopy of X-ray-selected clusters and the frequency of AGN in superclusters}

\author{E. Koulouridis\inst{1,2} \and B. Poggianti\inst{3} \and B. Altieri\inst{4} \and I. Valtchanov\inst{4} \and Y. Jaff\'e\inst{5} \and C. Adami\inst{6} \and A. Elyiv\inst{7,8} 
\and O. Melnyk\inst{9,10} \and S. Fotopoulou\inst{11} \and F. Gastaldello\inst{12} \and C. Horellou\inst{13} \and M. Pierre\inst{1} \and F. Pacaud\inst{14} \and M. Plionis\inst{15, 2, 16} 
\and T. Sadibekova\inst{1} \and J. Surdej\inst{17}}

\institute{Service d'Astrophysique AIM, DSM/IRFU/SAp, CEA Saclay, F-91191 Gif sur Yvette, France
\and Institute for Astronomy \& Astrophysics, Space Applications \&
  Remote Sensing, 
National Observatory of Athens, Palaia Penteli 15236, Athens, Greece
\and INAF - Astronomical Observatory of Padova,  Vicolo Osservatorio 5 - 35122 - Padova, Italy
\and European Space Astronomy Centre (ESA/ESAC), Operations Department, Villanueva de la Can\~{a}da (Madrid), Spain
\and Department of Astronomy, Universidad de Concepci\'on, Casilla 160-C, Concepci\'on, Chile
\and LAM, OAMP, Universit\'e Aix-Marseille, CNRS, P\^ole de l'\'Etoile, Site de Ch\~{a}teau Gombert, 38 rue Fr\'ed\'eric Joliot-Curie, 13388, Marseille 13 Cedex, France
\and Dipartimento di Fisica e Astronomia, Universit\`a di Bologna, Viale Berti
Pichat 6/2, I-40127  Bologna, Italy
\and Main Astronomical Observatory, Academy of Sciences of Ukraine, 27 Akademika
Zabolotnoho St., 03680 Kyiv, Ukraine 
\and Department of Physics, University of Zagreb, Bijenicka cesta 32, HR-10000 Zagreb, Croatia
\and Astronomical Observatory, Taras Shevchenko National University of Kyiv, 3
Observatorna St., 04053 Kyiv, Ukraine
\and Department of Astronomy, University of Geneva, ch. d'Ecogia 16, CH-1290 Versoix, Switzerland
\and INAF - IASF - Milano, Via Bassini 15, I-20133 Milano, Italy
\and Dept. of Earth and Space Sciences, Chalmers University of Technology, Onsala Space Observatory, SE-439 92 Onsala, Sweden
\and Argelander-Institut f\"ur Astronomie, University of Bonn, Auf dem H\"ugel 71, 53121 Bonn, Germany
\and Physics Department of Aristotle University of Thessaloniki,
University Campus, 54124, Thessaloniki, Greece
\and Instituto Nacional de Astrof\'{\i}sica Optica y Electr\'onica, Puebla,
C.P. 72840, M\'exico
\and Institut d'Astrophysique et de G\'eophysique, Universit\'e de Li\`ege, 4000
Li\`ege, Belgium}

\date{Received/Accepted}

\abstract{This article belongs to the first series of XXL publications. It presents multifibre spectroscopic observations of three 0.55 deg$^2$
fields in the XXL Survey, which were selected on the basis of their high density of X-ray-detected clusters. 
The observations were obtained with the AutoFib2+WYFFOS (AF2) wide-field fibre spectrograph 
mounted on the 4.2m William Herschel Telescope.}{The paper first describes the scientific rationale, the preparation, the data reduction, and the 
results of the observations, and then presents a study of active galactic nuclei (AGN) within three superclusters.}
{To determine the redshift of galaxy clusters and AGN, we assign high priority to a) the brightest cluster galaxies (BCGs), b) the most probable 
cluster galaxy candidates, and c) the optical counterparts of X-ray point-like sources. We use the outcome of the observations to study the projected (2D) and the spatial (3D) overdensity 
of AGN in three superclusters.}
{We obtained redshifts  for 455 galaxies in total, 56 of which are counterparts of X-ray point-like sources.
We were able to determine the redshift of the merging supercluster XLSSC-e, which consists of six individual clusters at $z\sim0.43,$ 
and we confirmed the redshift of supercluster XLSSC-d at $z\sim0.3$. More importantly, we discovered a new supercluster, XLSSC-f, 
that comprises three galaxy clusters also at $z\sim0.3$. 
We find a significant 2D overdensity of X-ray point-like sources only around 
the supercluster XLSSC-f. This result is also supported by the spatial (3D) analysis of XLSSC-f, where we find four AGN with compatible 
spectroscopic redshifts and possibly one more with compatible photometric redshift.
In addition, we find two AGN (3D analysis) at the redshift of XLSSC-e, but no AGN in XLSSC-d. Comparing these findings with the optical
galaxy overdensity we conclude that the total number of AGN in the area of the three superclusters significantly exceeds the field expectations.
All of the  AGN found have luminosities below $7\times10^{42}$erg s$^{-1}$.}
{The difference in the AGN frequency between the three superclusters cannot be explained 
by the present study because of small number statistics.  
Further analysis of a larger number of superclusters within the 50 deg$^2$ of the XXL is needed 
before any conclusions on the effect of the supercluster environment on AGN can be reached.}

\authorrunning{E. Koulouridis et al.}
\titlerunning{The XXL survey - XII}

\maketitle

\section{Introduction}

As structures grow hierarchically, galaxies are accreted by progressively more massive dark matter halos, and  the majority of
galaxies end up in clusters (Eke et al. 2004; Calvi et al. 2011). Clusters are therefore the predominant environment of galaxies and
can play a very important role in establishing galaxy properties.

Although there is no explicit classification, galaxy concentrations with more than 50
members and more massive than $10^{14}M_{\sun}$ are defined as galaxy clusters. Less massive aggregations with less than 
50 galaxies are called galaxy groups. 
We note that according to the above classification most of the extended X-ray sources in the current study are clusters.

Clusters and groups are usually identified by optical and infrared surveys 
as concentrations of red-sequence galaxies (e.g. Gladders \& Yee 2000; Koester et al.
2007a; Hao et al. 2010; Rykoff et al. 2014; Bleem et al. 2015) or galaxy overdensities in photometric redshift space 
(e.g. Wen, Han \& Liu 2009, 2012; Szabo et al. 2011) and they are confirmed by follow-up spectroscopy. 
They can also be identified by X-ray observations as extended sources, 
unambiguously testifying the presence of hot gas trapped in the potential well of a virialised system (e.g. Pierre et al. 2004; 
Pacaud et al. 2007; Pierre et al. submitted, hereafter XXL paper I). X-ray selected cluster samples 
are  rarer and smaller than optically selected
ones, and deep X-ray observations are required to probe a significant range of halo masses.

The properties of galaxy populations in groups and clusters vary enormously. At low redshift, it is well known that some galaxy groups are dominated
by early-type, passively evolving galaxies, similarly to clusters, while others have a galaxy population resembling that of the field, mostly
composed of late-type, star-forming galaxies (Zabludoff \& Mulchaey 1998). Recent studies of optically selected clusters at intermediate
redshifts have found a similar variety. Surveys like EDisCS (Poggianti et al. 2006, 2009), zCOSMOS (Iovino et al. 2010), and CNOC2
(Wilman et al. 2005, 2008) find that cluster galaxies  differ significantly from galaxies that reside in lower mass halos in the field, but with
a wide range of properties at a given cluster velocity dispersion. Whether this variety originates from the difference between
virialised clusters and clusters in formation or from unbound galaxy associations is still an open question, especially given the broad spread
in galaxy properties observed in the currently small X-ray selected samples (Jeltema et al. 2007; Urquhart et al. 2010).

The effect of the group and cluster environment on the activity of the central supermassive black hole (SMBH) of galaxies
and vice versa is still fairly undetermined, but nevertheless crucial. 
Galaxy clusters represent one end of the 
density spectrum in our universe, and as such they are an ideal place to
investigate the effect of the dense environment in the triggering of active galactic nuclei (AGN), especially since 
an excessive number of X-ray point-like sources are undoubtedly found there
(e.g. Cappi et al. 2001; Molnar et al. 2002; Johnson et al. 2003; D'Elia et al.
2004; Cappelluti 2005; Gilmour et al. 2009). Specifically, for the XMM-LSS field, 60\% 
of X-ray-selected AGN reside in the overdense regions of group-like environment (Melnyk et al. 2013). We note that
AGN can be used as cosmological probes to trace the large-scale structure at high redshifts (e.g. Einasto et al. 2014), and thus
the study of the AGN frequency-to-density relation is essential.

Theoretically, the feeding of the black hole can only be achieved by means of a 
non-axisymmetric perturbation that induces mass inflow. This kind of 
perturbation can occur in interactions and merging between two galaxies,
which results in the feeding of the black hole and the activation of the AGN
phase (e.g. Umemura 1998; Kawakatu et al. 2006; Koulouridis et al. 2006a,
2006b, 2013; Koulouridis 2014; Ellison et al. 2011;
Silverman et al. 2011; Villforth et al. 2012; 
Hopkins \& Quataert 2011). Thus, the cluster environment, where 
the concentration of galaxies is very high relative to the field, would also
seem favourable to
AGN. However, the rather extreme 
conditions within the gravitational potential of a galaxy cluster can work in
the opposite direction as well. The ram pressure from the intracluster medium
(ICM) is probably able to strip or evaporate the cold gas reservoir of galaxies
(Gunn \& Gott 1972; Cowie \& Songaila 1977; Giovanelli \& Haynes 1985; Chung et al. 2009; Jaff{\'e} et al. 2015) and can strongly affect 
the fueling of the AGN. 
Other studies, however,
have argued that ram pressure stripping cannot be as effective
in transforming blue-sequence galaxies to red
(e.g. Larson et al. 1980; Balogh et al. 2000, 2002;
Bekki et al. 2002; van den Bosch et al. 2008; Wetzel et al. 2012), especially
in galaxy groups where other processes are taking place as well. 
In addition, possible prevention of 
accretion of gas from the halo into cluster or group galaxies (``strangulation'';
e.g. Larson et al. 1980; Bekki et al. 2002; Tanaka et al. 2004) may, in fact, suppress 
AGN activity. 

When using only optically selected AGN, the results on the AGN frequency within galaxy clusters 
remain inconclusive.
Early studies reported that AGN are less 
frequent in galaxy clusters than in the field (Osterbrock 1960; Gisler
1978; Dressler, Thompson \& Schectman 1985) and more recent studies
support this suggestion (Kauffmann et al. 2004; Popesso \& Biviano
2006; von der Linden et al. 2010; Pimbblet et al. 2013). Other studies,
however, have found no differences between cluster and field galaxies (e.g. Miller
et al. 2003). 

In contrast to optically selected AGN, radio-loud AGN seem to be more clustered than any
other type of galaxy (Hart, Stocke \& Hallman 2009) and are often
associated with BCGs (brightest cluster galaxies) (e.g. Best 2004; Best
et al. 2007). Nevertheless, Best et al. (2005) showed that radio-loud AGN 
with the strongest optical emission lines avoid the densest regions, a fact that
implies a certain connection between the environment and the accretion rate onto the 
SMBH.

Undoubtedly, the best way to detect active galaxies is through X-ray
observations (e.g. Brandt \& Alexander 2010). 
During the previous decade, spectroscopic studies of X-ray point-like 
sources in rich galaxy clusters have concluded that low-X-ray-luminosity AGN ($<3\times 10^{42}$erg s$^{-1}$) 
are equally present in cluster and field environments (e.g. Martini et
al. 2007; Haggard et al. 2010), although most of them presented no optical AGN spectrum (e.g. Martini et
al. 2002, 2006; Davis et al. 2003). Nevertheless, luminous AGN were rarely found in clusters 
(Kauffmann et al. 2004; Popesso \& Biviano 2006).
More recent 
studies also reported a significant lack of AGN in rich galaxy clusters by comparing X-ray to optical
data. Koulouridis \& Plionis (2010) demonstrated the
suppression of X-ray-selected AGN in 16 rich Abell clusters (Abell 1958) by comparing
the X-ray point source overdensity to the optical galaxy
overdensity. Ehlert et al. (2013; 2014) found that the X-ray AGN fraction
in the central regions of 42 of the most massive clusters known
is about three times lower than the field value using the same
technique, while in their most recent study (Ehlert et al. 2015) they argue that 
galaxy mergers may be an important contributor to the cluster AGN population. 
More importantly, from the complete
spectroscopy of their X-ray point-like source sample, Haines et al. (2012) concluded
that X-ray AGN
found in massive clusters are an in-falling population and confirm the suppression in 
the inner regions of rich
clusters. On the other hand, Martini et al. (2013) argue that this
trend is not confirmed for a sample of high-redshift clusters
($1.0<z<1.5$).
Finally, an indirect way to address the issue is by
clustering analyses, but  these results also remain 
inconclusive (see relevant discussion in Haines et al. 2012 \S5.2).
 
The majority of
the above studies  deal with AGN within massive clusters, while the presence of AGN in less massive or even more 
massive formations has been very poorly studied. In a scenario in which AGN are suppressed
by the strong gravitational potential of massive clusters (through gas
stripping, strangulation, tidal stripping, evaporation, high velocity-dispersion,
etc.), one would expect the AGN presence to rise in shallower gravitational potentials 
(see Arnold et al. 2009; Gavazzi et al. 2011; Bitsakis et al. 2015) and be completely
nullified within the deepest ones. 
In Koulouridis et al. (2014), we investigated
the AGN presence in two samples of poor and moderate clusters and found evidence of this 
anti-correlation. 
Interestingly, in merging or 
actively growing clusters the high incidence of galaxy mergers can potentially enhance the number of AGN,
while at the same time, shock waves may also enhance the ram pressure stripping intensity (Vijayaraghavan \& Ricker 2013; 
Jaff{\'e} et al. in prep.).

In the current study we investigate the most extreme massive formations in the Universe, superclusters.
They typically consist of three to ten clusters spanning as many as 150 $h^{-1}$Mpc and are without 
sharply defined boundaries (e.g. Chon et al. 2014, Pearson 2015).
The superclusters can vary widely in size, containing from a few small groups of the order of
$10^{13} - 10^{14} M_{\sun}$ (e.g. Einasto et al. 2011; Chon et al. 2014) up to many massive clusters.
We note, however, that the mass density, averaged on the supercluster scale, is smaller than in clusters.
They are already decoupled from the Hubble flow, but
not yet virialised;  the time it takes a randomly moving galaxy to traverse 
the long axis of a supercluster is typically comparable to the age of the universe. 
They also appear to be interconnected, but the boundaries between them are poorly defined.
At these large scales the dynamical evolution proceeds at a slow rate and superclusters 
reflect the initial conditions of their
formation. Therefore, they are important sites where we can directly 
witness the evolution of structure formation and mass assembly.
 
With its depth, uniform coverage, and well-defined selection function,
the XXL Survey (The Ultimate XMM-Newton Survey, XXL paper I) is making a unique contribution to the study of distant
clusters. In addition, its two 5x5 deg$^2$ fields are essential to the study of AGN in the cluster environment. Clusters can be 
very extended, of the order of a few Mpc, and AGN may preferentially reside even further out in their outskirts (e.g. Fassbender et al. 2012; Haines et al. 2012;
Koulouridis et al. 2014). 
More than half of the detected extended sources are 1-3 keV clusters in the $0.2<z<0.5$ range
(Fig. 1), they cover an estimated mass range $10^{12.8} - 10^{14.5} M_{\sun}$, and are the subject of our spectroscopic follow-up campaign.

In the first part of the current paper (\S2 and \S3), we present the preparation, the data reduction, and the 
results of the  William Herschel Telescope (WHT) observations. In the second part (\S4) we investigate the AGN frequency within
the three superclusters. The results for cluster galaxies and the related
  spectroscopic catalogues will be presented in a subsequent paper. Throughout
this paper we use $H_0=70$ km s$^{-1}$ Mpc$^{-1}$, $\Omega_m=0.28$, and
$\Omega_{\Lambda}=0.72$.

\section{Data description}
\subsection{The XXL Survey}

The XXL Survey is the largest XMM project approved to date ($>$6 Msec), 
surveying two $\sim$ 5x5 deg$^2$ fields at a  depth of $\sim5\times10^{-15}$ erg sec$^{-1}$ cm$^{-2}$ in the [0.5-2] keV soft X-ray band\footnote[1]{
The XXM-Newton observation IDs used in the current study:\\
Field-1:\\ 
0677670135, 0677670136, 0677680101, 0677680131, 0677681101\\
Field-2:\\ 
0651170501, 0651170601, 0655343860, 0677650132, 
0677650133, \\0677650134, 0677660101, 0677660201,
0677660231, 0677660232, \\0677660233, 0677670133,
0677670134, 0677670135, 0742430101         \\
Field-3:\\ 
0109520201, 0109520301, 0111110101, 0111110201,
0111110701, \\0112680101, 0112680401, 0112681001,
0112681301, 0677580131,\\ 0677580132, 0677590131,
0677590132, 0677590133} (completeness limit for the point-like sources).
The XXL observations have been completed and processed. 
To date some 450 new galaxy clusters have been detected out to redshift $z\sim2$ as well as more than 10000 AGN 
out to $z\sim4$. The main goal of the project is to constrain the Dark Energy equation 
of state using clusters of galaxies. This survey will also have lasting legacy value for cluster scaling laws 
and studies of galaxy clusters, AGN, and X-ray background. The northern field (XXL-N), which we use in the current study, is also covered in other 
wavelengths, e.g. the Canada-France-Hawaii Telescope Legacy Survey (CFHTLS-optical), Spitzer Space 
Telescope (SST-infrared), the UKIRT Infrared Deep Sky Survey (UKIDSS) and the Galaxy Evolution Explorer (GALEX-Ultraviolet).

\subsection{Spectroscopic target and supercluster selection}
The three fields observed in this work (see Table 1) were chosen on the basis of the high number of X-ray
clusters, containing a total of 25 X-ray groups/clusters in the redshift range that 
we are targeting, i.e. $0.2<z<0.5$.
In order of priority, we targeted a) all the BCGs,
b) cluster galaxy candidates selected on the basis of projected distance to the cluster X-ray position ($<500h^{-1} kpc$ and $19<r_{SDSS}<21$),
c) optical counterparts of X-ray point-like sources (mostly AGN), and finally d) any other galaxy in the targeted 
redshift range according to their photometric redshift.

Superclusters are defined as concentrations of clusters that trace a second-order clustering hierarchy of galaxies, and they are the
largest structures observed. In the current study we identify superclusters as concentrations of at least three clusters 
at a close redshift separation within $25'$ radius, given the limited field of view (FoV) of the WHT. 
Our three observed fields include a  total of three superclusters (see Table 2). 

Pacaud et al. (submitted, hereafter  XXL paper II) base their selection on a different methodology because of the different sample 
(the 100 brightest clusters, hereafter XXL-100-GC\footnote[2]{XXL-100-GC data are available in computer readable form via the
XXL Master Catalogue browser http://cosmosdb.iasf-milano.inaf.it/XXL, and via the XMM XXL DataBase http://xmm-lss.in2p3.fr})
and the different area (the full XXL Survey). 
According to XXL paper II a supercluster must include a close pair of clusters ($D<8 h^{-1}$ Mpc)
and at least a third cluster within 20 $h^{-1}$ Mpc of the pair. The 
above selection requires all three clusters of the starting triplet to be members of the XXL-100-GC. Then all clusters within 35 $h^{-1}$ Mpc, independent of brightness, are 
considered  supercluster members.  They finally report five superclusters, XLSSC-a to -e.

Two of them are in common with the current paper,
i.e. Field-1 includes XLSSC-e and Field-2 a part of XLSSC-d. The latter comprises seven X-ray detected clusters in XXL paper II, but the WHT FoV 
includes only a close bright pair and one more fainter cluster\footnote[3]{Because of the slightly higher average redshift of the three clusters of XLSSC-d in the current paper ($z$=0.298) 
than of the seven clusters in XXL paper II ($z$=0.294), the two papers report slightly different redshift (0.30 and 0.29, respectively).}. 
In Field-3 we discover a supercluster that satisfies the first criterion 
of XXL paper II of having a close pair of clusters that belong to the XXL-100-GC sample, but the third member is fainter.
We name this supercluster XLSSC-f.

In addition to the discovery of the XLSSC-f supercluster, we also publish a new XXL cluster,
namely XLSSC 117. We list some basic properties of the new cluster in Table 2. 

\section{Multifibre optical spectroscopy}

\subsection{Target preparation}

For the preparation of the observations 
we executed the software {\tt af2-configure}, available in the Isaac Newton Group of Telescopes (ING) website\footnote[4]{http://www.ing.iac.es}. 
It is designed to create the mapping between the objects and the fibres during
a particular spectrograph exposure. It uses an input file with the coordinates ($\alpha, \delta$) of the objects, creates a fibre-to-object 
mapping using one of two currently available placement algorithms, and then allows the user to edit the fibre locations interactively.
In the input file the user should also assign priority to all objects. High priority should  be assigned 
to fiducial stars, since it is essential to allocate approximately eight fiducial fibres, scattered homogeneously 
in the field, to accurately align the science fibres.
The placement algorithms search for the best combination of position angle of the spectrograph on the sky
and targets in the fibres that maximise the sum of object priorities. 

The fibres are positioned by {\tt af2-configure} within a FoV of 1 degree in diameter, but we 
manually limited our targets
within the central 25 arcmin radius to avoid the
effects of vignetting. We tried 
to maximise the number of
fibres allocated on galaxies, but typically also placed  20-30
fibres on the sky for sky subtraction purposes. Within each field our targets were divided into bright ($19<r_{SDSS}<20.5$) and faint
($20.5<r_{SDSS}<21$) and we prepared two fibre configurations for the bright sources and one for the faint. We
allocated an average of $\sim$100 sources 
per configuration, plus sky fibres and 
fiducial stars.

\subsection{Observations}

We observed the three fields with the 4.2m WHT during six nights in 2013. More details about the observations are listed in Table 1.
We conducted multifibre medium resolution spectroscopy with the AutoFib2+WYFFOS (AF2)
wide-field multifibre spectrograph. The
AF2 contains 150 science fibres of 1.6 arcsec diameter 
and 10 fiducial bundles for acquisition and guiding. 
At the prime focus, the fibres are placed onto a field
plate by a robot positioner at user-defined sky coordinates (see \S3.1).

\begin{table}
\caption{WHT observations} 

\tabcolsep 3 pt
\def\arraystretch{1.3}
\begin{tabular}{lcllrc} \hline
\\
Date & Time U.T. & Name  & Config.& Exp. & Seeing \\
  (1)    &  (2) &  (3)&  (4)     &  (5)& (6)  \\
\hline
2013 Oct 29 &22:00 - 00:50 & Field-1 & 1st Bright & $150$  & $1.5''-2.4''$\\   
2013 Oct 29 &01:42 - 04:45 & Field-1 & 2nd Bright & $150$  & $1.5''-2.4''$\\
2013 Oct 30 &22:00 - 00:15 & Field-3 & 1st Bright & $120$  & $<1.5''$\\
2013 Oct 30 &00:50 - 04:30 & Field-1 & Faint & $200$  & $<1.5''$\\
2013 Nov 07 &22:00 - 02:10 & Field-3 & Faint & $240$  & $<1.5''$\\
2013 Nov 08 &21:00 - 00:20 & Field-3 & 2nd Bright & $180$  & $>1.5''$\\
2013 Nov 08 &01:20 - 02:20 & Field-2 & 1st Bright & $60$  & $>1.5''$\\
2013 Nov 09 &23:00 - 02:30 & Field-2 & 2nd Bright & $180$  & $>2.0''$\\
2013 Nov 10 &22:00 - 23:30 & Field-2 & 1st Bright & $120$  & $>1.5''$\\
2013 Nov 08 &00:00 - 03:30 & Field-2 & Faint & $180$  & $0.5''-0.9''$\\
\hline
\end{tabular}
\tablefoot{{\em (1)} Date of observation, {\em (2)} starting and ending U.T., {\em (3)} name of the observed WHT field, {\em (4)} target selection: ``Bright'' for
targets $19<m_r<20.5$ and ``Faint'' for $20.5<m_r<21$, {\em (5)} exposure time in minutes, {\em (6)} seeing during the observation.} 
\end{table}

We used the R600B grating with the new default detector Red+4. It is an e2v 231-84 4k$\times$4k, red-sensitive, fringe-suppression CCD
with a mosaic of 4096$\times$4112 pixels, 15$\mu$m each. We used a 2$\times$2 binning of the CCD pixels and we obtained a spectral
resolution of $\sim$4.4 \AA. The spectra were centred at wavelength $\sim$5400 \AA, and covered the range 3800 to 7000 \AA. The spectra 
of He and Ne lamps were used for the wavelength calibration.
 
The bright configurations were observed for 2 or 2.5 hours each (depending on the seeing), 
and the faint for 3-4 hours each. We were able to observe nine fibre configurations. In total the run yielded $\sim$900 spectra.

\subsection{Data reduction}

Data were reduced using the AF2 data reduction pipeline v1.02\footnote[5]{a newer version of the pipeline (v3.0) can be downloaded from 
http://www.ing.iac.es/astronomy/instruments/af2/reduction.html}. 
The pipeline is written in IDL and is able to perform
data reduction, including fibre-to-fibre sensitivity corrections
and optimal extraction of the individual spectra. 
Below we describe briefly the calibration and extraction modules of 
the pipeline, but more details can be found in the pipeline 
manual distributed online by the ING.

The first steps of the pipeline include master bias correction, tracing
of the fibres, flat-field correction, masking of bad pixel 
in the science data, and wavelength calibration. In more detail:

\begin{enumerate}
 \item {\tt BIAS} module: At least ten bias files are used each night to debias all raw data images. 
 The average signal level in the overscan regions 
 is used to correct for any change in the bias level over time.
\item {\tt MASK} module: The module produces a mask file of the CCD pixels where the dark
current exceeds a user-specified level. It also displays a plot of the fraction of masked pixels 
versus the cut-off level and an image of the produced file.
\item {\tt FLAT} module: At least ten twilight sky or internal flats were used to perform
the flat-field correction each night. Individual flats are scaled according to their mean value
before calculating their total median value.
\item {\tt CIRC} module: This module uses a flat file to trace the x-pixel
position of the centre of the spectral line of each active fibre as a function
of y-pixel. In the pipeline version used for the reduction in the current paper, the module crashed if the low-signal area of the CCD 
(the blue part of the spectrum) was not trimmed. The new version of the pipeline, however, does not present this problem and 
the user can analyse the CCD in its full length.  
\item {\tt ARC} \& {\tt ATLAS} module: The module extracts the
lamp spectrum as a function of y-pixel position and uses this intermediate spectrum to
determine the wavelength calibration. In our case two arc files are used to reference lines simultaneously, 
one from the helium lamp for the blue part of the spectrum, and one from the neon lamp for the red part of 
the spectrum. The {\tt ARC} module 
identifies the approximate y-pixel location and exact wavelengths for a set of well-separated unsaturated 
lines in the arc spectra and finds the precise position of the peaks by fitting Gaussian profiles 
to each one. It uses a predefined table of emission line data, but in combination with the {\tt ATLAS} module 
the selection and confirmation of the lamp lines is performed interactively.
\end{enumerate}

The extraction of the spectrum by the pipeline is done in two additional steps: 

\begin{enumerate}
  \item {\tt STAR} module: The module first extracts
the science spectra of designated targets and sky-allocated fibres and then processes the
intermediate spectra to produce sky subtracted output spectra on a common
wavelength base. The median sky spectrum is calculated within the {\tt STAR} module. There are four different 
options for the calculation, but in our case we selected the one where the median 
sky is scaled and the output spectrum is masked over sky lines. 
  \item {\tt MEDAN} module: This module evaluates the median spectra for
each fibre by combining all available science exposures. Spectra are normalised to their mean value before the
median is calculated.
 \end{enumerate}

Finally, the flux calibration, which is not included in the pipeline, is performed with {\sc IRAF} using {\tt STANDARD}, {\tt SENSFUNC}, and
{\tt CALIBRATE} tasks. Given the wide magnitude range covered, the spectra have a wide range of S/N.

\subsubsection{Galaxy redshifts}

Redshifts were obtained from visual inspection of all spectra by two of
the authors (BP and CA), using the {\sc IRAF} package {\sc RVSAO} independently and
iterating on doubtful cases.

The overall success rate (number of redshifts/number of spectra) was 60\%, 
ranging from 84\% for the best seeing configurations to
 the lowest 30\% for the worst ones.

The probability that each redshift was correct was estimated on the basis of
the number and quality of spectral features (lines in emission or absorption, D4000) identified in the spectrum. 
If a large number of lines were identified without wavelength offsets the probability was set to 99\%.
As the number of identified spectral features decreased, the assigned probability
decreased;   75\% percent redshifts had at least two or three secure lines.
 
 In total we obtained 455 good-quality redshifts (of which
   172/147/110 have a 75\%/95\%/99\% chance of being correct, and
   26 are based on a single emission line but with very reliable identification).
 Ninety other redshifts were more uncertain and were not used for any of our purposes, but were recorded
 for future reference. We note that our target sample was not contaminated by stars.

 In more detail we obtained:
\begin{enumerate}
\item 9 BCG (brightest cluster galaxy) redshifts. These provide us with nine new cluster redshifts 
around $z$=0.3-0.5.
More importantly these redshifts are too high to be obtained by GAMA spectroscopy (Driver et al. 2011; Liske et
al. 2015), which covers the XXL Survey, but  is much shallower.

\item 82 and 160 cluster galaxy candidates that lie within 0.5 and 1 $h^{-1}$ Mpc of the cluster centres, respectively.
 We will use these galaxies for a more precise determination of the cluster redshifts and for the study of 
 the cluster properties.
   
In addition, we obtained the redshifts of another 
 148 galaxies that lie more than 1 $h^{-1}$ Mpc  from the closest extended X-ray source.

 \item 56 AGN redshifts that are used  
 in the second part of the current paper, in combination with spectroscopic results from other surveys.
 \end{enumerate}

In Fig.1 we compare our spectroscopic redshifts
with the CFHTLS photometric redshifts (Ilbert et al. 2006 and Coupon et al. 2009). 
WHT spectroscopy is clearly useful in the redshift range of interest ($0.2<z<0.5$).

\begin{figure}
\resizebox{8cm}{8cm}{\includegraphics{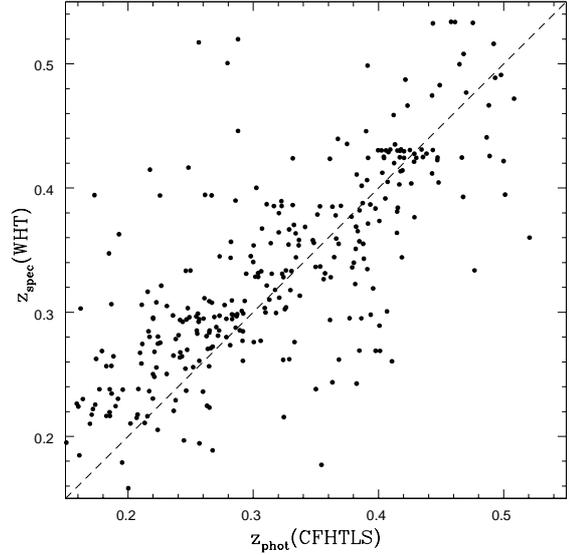}}
\caption{Comparison of the galaxy spectroscopic redshifts obtained during the two runs with the William Herschel Telescope
with the CFHTLS photometric redshifts.}
\end{figure}

\begin{table*}
\centering
\caption{Superclusters} 
\tabcolsep 5 pt
\def\arraystretch{1.3}
\begin{tabular}{ccc|ccc|lcccccc} \hline
Obs. Field  & RA  & Dec & Supercluster & RA & Dec& cluster ID &  $z_{spec}$ &  $z_{mean}$& $T_{300kpc}$ & $M_{500,MT}$&$r_{500,MT}$&ref.\\  
  (1)    &  (2)&  (3) &  (4)&  (5)     &  (6)& (7) &(8) &(9)&(10) &(11)&(12)&(13) \\
\hline
Field-1 & 32.60 & -6.30 & XLSSC-e  & 32.87 & -6.20  & XLSSC 081 &  0.432 & &  1.7$^{+0.3}_{-0.2}$ & 0.7 &0.55&5\\   
     &  &    &  &    &  &XLSSC 082$\dagger$  & 0.427& & 3.9$^{+1.7}_{-0.6}$  & 2.9&0.88&5\\
      &  &   &  &    &  &{\bf XLSSC 083}$\dagger$  & 0.430& & 4.8$^{+1.2}_{-0.9}$  & 4.1&0.99&4\\
     &  &    &  &    &  &{\bf XLSSC 084}$\dagger$  & 0.430& 0.429& 4.5$^{+2.3}_{-1.5}$  & 3.7&0.96&4\\
      &  &   &  &    &  &{\bf XLSSC 085}$\dagger$  & 0.428& & 4.8$^{+2.0}_{-1.0}$  & 4.1&0.99&4\\
      &  &   &  &    &  &XLSSC 086$\dagger$  & 0.424& & 2.6$^{+1.2}_{-0.6}$  & 1.5&0.70&5\\
         \hline
Field-2  &  36.93 & -4.70  &XLSSC-d & 37.22 & -5.05  &XLSSC 013 &  0.307 & & 1.6$^{+0.3}_{-0.1}$ & 0.7 & 0.57&1\\
       &  &   & &    &  &{\bf XLSSC 022}  & 0.293& 0.298& 2.1$^{+0.1}_{-0.1}$  & 1.1 &0.68&2\\
       &  &   & &    &  &{\bf XLSSC 027} & 0.295& & 2.7$^{+0.4}_{-0.3}$  & 1.7 &0.77&3\\
         \hline
Field-3  &  33.12 & -5.82   &  XLSSC-f  & 33.12 & -5.82  & {\bf XLSSC 098} &  0.297 & & 2.9$^{+1.0}_{-0.6}$  & 1.9 &0.81&4\\
        &  &  & &    &  &{\bf XLSSC 111} & 0.299 &0.298& 4.5$^{+0.6}_{-0.5}$  & 4.0&1.02&4\\
       &  &   & &    &  &XLSSC 117 & 0.298 &&  3.3$^{+0.8}_{-0.7}$  & 2.4  &0.86&6\\
         
\hline
\end{tabular}
\tablefoot{{\em (1)} Observed field name, {\em (2)-(3)} field coordinates in the J2000 system, {\em (4)} supercluster name, {\em (5)-(6)}, supercluster coordinates 
in the J2000 system, as published in XXL paper II for XLSSC-d and -e, 
{\em (7)} original cluster name in the XXL database, members of the XXL-100-GC sample in bold,
{\em (8)} spectroscopic redshift, {\em (9)} mean supercluster redshift,
{\em (10)} X-ray temperature in keV within an aperture of 300 kpc measured in Giles at al. (submitted, XXL paper III) for the members of the XXL-100-GC sample, 
{\em (11)} cluster mass in $10^{14}M_{\sun}$, calculated from the $M_{500,MT}-T_{300kpc}$ scaling relation of Lieu et al. (submitted, hereafter XXL paper IV), 
{\em (12)} overdensity radius with respect to the critical density in Mpc, calculated from the $M_{500,MT}-T_{300kpc}$ scaling relation of XXL paper IV, 
{\em (13)} reference to the first X-ray detection as a cluster.\newline
{\bf References.} (1) Willis et al. (2005); (2) Pierre et al. (2006); (3) Pacaud et al. (2007); (4) XXL paper II; (5) XXL paper VII; (6) this work.    
\newline
$\dagger$ The spectroscopic redshift of the cluster was initially determined by the observations presented in the current paper.}
\end{table*}

\section{AGN in superclusters}         

In the following sections we present a study of the AGN frequency in the three observed supercusters. 
The supercluster in Field-1 (XLSSC-e) is very different from the ones found in the other two fields, i.e. all five clusters found at z$\sim$0.43 
are located within a circle of $4'$ radius (1.3 $h^{-1}$ Mpc) and a sixth but uncertain member within $10'$.
A more detailed analysis of the supercluster and its BCGs is presented in Pompei et al. (submitted, hereafter XXL paper VII). Baran et al. (submitted, hereafter XXL paper IX)
identified several overdensities via a Voronoi 
Tessellation analysis of optically detected galaxies and presented new radio observations. 
Therefore, we consider the supercluster in Field-1 to be a merging supercluster in a tight configuration. 

On the contrary, in the FoV of the WHT (6.5 $h^{-1}$Mpc radius at $z=0.3$), the other two structures 
include only three members each in a much looser configuration, although both include a very close cluster pair ($<1$ $h^{-1}$Mpc). 
Therefore,
the superclusters in Field-2 (XLSSC-d) and Field-3 (XLSSC-f) seem very similar (see Fig. 2).
However, further investigation outside the WHT FoV reveals that these two superclusters are intrinsically different, i.e. the structure found in Field-2 is only part 
of a larger formation comprising seven X-ray detected clusters within 35 $h^{-1}$Mpc (XXL paper II), 
while the three clusters found in Field-3 are not related to any 
significant overdensity in the region. Nevertheless, the total mass of the latter is larger by a factor of 2.5. 

Because of the above differences and its higher redshift, 
we studied AGN in XLSSC-e within a $10'$ radius around its five confirmed clusters,
while for the other two superclusters we  used the full $25'$ FoV of the WHT. 
In general, superclusters are not virialised and there is no explicit 
definition of their centre. For XLSSC-e we chose 
a position approximately in the middle of the formation to be the centre, while for XLSSC-d and -f
we used the centre of the WHT FoV. From the X-ray images in Fig. 2 (right panels) 
it is apparent that the above choices are reasonable.
 
\subsection{Methodology}
We assessed the enhancement or the suppression of AGN presence within the three superclusters by
analysing both the 3D (spatial) and the 2D (projected) overdensity of X-ray point sources. We chose to analyse  the 2D case as well since we lacked 
complete spectroscopy for all the X-ray point sources. However, we had to take into account 
that the 2D case is hampered by a variety of systematic effects, related for example to flux-boosting due to lensing 
(see discussion in Koulouridis et al. 2014). For the statistical evaluation of our results we used the confidence limits for small numbers
of events in astrophysical data, based on Poisson statistics (Gehrels 1986). 

To assess the 2D and 3D overdensity of AGN in the three superclusters we adopted a common lower luminosity limit for the X-ray point-like sources. 
We find that for the two superclusters at $z=0.3$ a luminosity limit of $L_{(0.5-2{\rm\; keV})}= 2.7\times 10^{42}$erg s$^{-1}$,
which corresponds to $f_{(0.5-2{\rm\; keV})}\sim 1.0\times10^{-14}$ erg s$^{-1}$ cm$^{-2}$, combines both the inclusion of low-luminosity 
AGN and a relatively high completeness of spectroscopic redshifts. For the supercluster XLSSC-e at $z=0.43$, this luminosity limit corresponds 
to a flux limit of $f_{(0.5-2{\rm\; keV})}=4.5\times10^{-15}$ erg s$^{-1}$ cm$^{-2}$. We will show the importance of having a common 
luminosity rather than a flux limit for fields at different redshifts.

\subsubsection{Projected overdensity of X-ray point-like sources}
In a given area, the projected overdensity of X-ray AGN is estimated according to
\begin{equation}
\;\;\;\;\;\;\;\;\;\;\;\;\;\;\;\;\;\;\;\;\;\;\;\delta_x=\frac{N_x}{N_{\rm exp}}-1,
\end{equation}
where $N_x$ is the
number of X-ray point-like sources detected in the area and $N_{\rm exp}$
is the expected number according to the $\log N-\log S$ of the XXL northern field (Elyiv et al. in prep) within the same
area. We note that the soft-band $\log
N - \log S$ used in the current study is lower than 
those of the 2XMM (Ebrero et al. 2009) and COSMOS (Cappelluti et al. 2009) surveys (with deviations not
exceeding the $2 - 3\sigma$ Poisson level), but in excellent agreement with those of the XMM Medium Deep Survey (XMDS, Chiappetti et al.
2005).

To calculate the value of $N_x$, we identified all point-like sources located
within five radial annuli between $n$ and
$(n+1) r$ around the centre of each field, where $n=$0,1,2,..5, $r=2'$  for XLSSC-e (higher redshift and more compact) and $r=5'$ for XLSSC-d and XLSSC-f. 
The large contiguous area of the XXL Survey allowed us to expand
our search for X-ray AGN at large radii.

To calculate the expected number $N_{\rm exp}$ of X-ray sources in the
field, we followed the procedure described below,
considering each time the same area of the detector and 
the same characteristics of the actual observation: 
\begin{enumerate} 
\item From the $\log N-\log S$ we derived the total number ($N_f$) of expected sources in the area per flux bin.
\item We considered $1000\times N_f$ sources with random fluxes within
  the flux range of each bin and random positions within the area of
  interest.
\item We derived the probability $P_i$ that the source $N_{fi}$ is
  actually detected in the specific area of the detector. The probability
  is a function of the off-axis position (vignetting), background, and 
  exposure time (Elyiv et al. 2012).
\item We calculated the sum $\sum_{j=n}^{nbin}\sum_{i=1}^{1000N_f}
  N_{fi}\times P_i/1000$, which gives  the total number, $N_{\rm exp}$, 
of expected X-ray sources that have fluxes above the respective value
of the $n^{th}$ bin of the $\log N-\log S$, where the total number of bins, nbin, is
160.
\end{enumerate}

\subsubsection{Spatial overdensity of X-ray point-like sources}
\begin{figure*}
\resizebox{18.5cm}{22cm}{\includegraphics{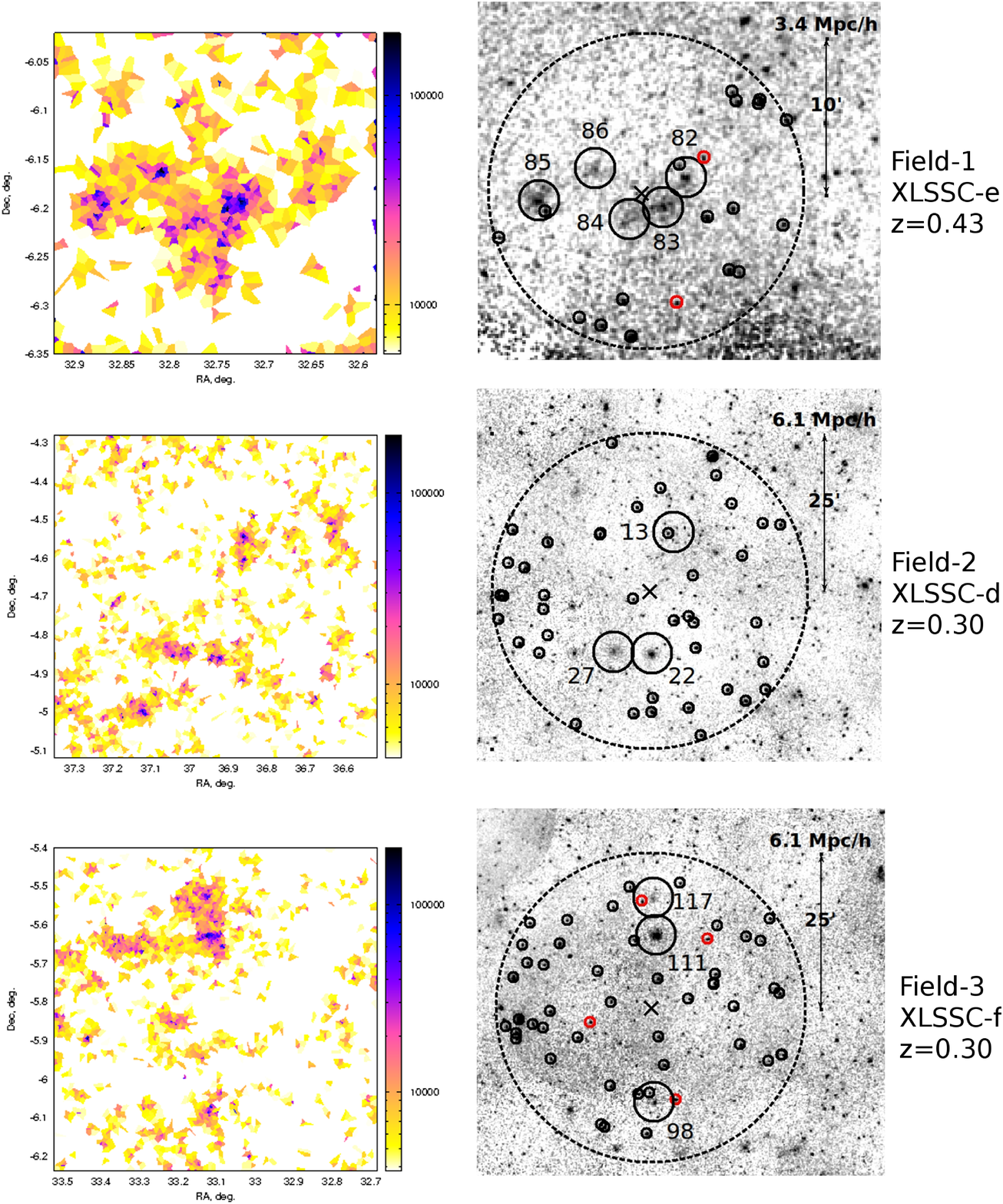}}
\caption{Voronoi tessellations (VT) and X-ray images of the three superclusters (XLSSC-e: top, XLSSC-d: middle, XLSSC-f: bottom). 
Left panels: Voronoi tessellation
using CFHTLS photometric redshift data. All galaxies within $\delta z=\pm0.05$ of the supercluster redshift are included. 
The side bar illustrates the colour-coding of the galaxy number-density. 
Only areas above the average number-density are coloured in the plots.  
Right panels: the corresponding X-ray maps overplotted with
the positions of the studied areas (dashed circles) and of the X-ray detected clusters (large black circles with XLSSC ID numbers). 
The small black circles denote X-ray point-like sources above the luminosity limit
($L_{(0.5-2{\rm\; keV})}>2.7\times 10^{42}$ erg s$^{-1}$), while the red circles denote the ones with spectroscopic redshift consistent
with the supercluster (within $2\times\delta z$, see \S4.1.2). The supercluster centres are marked with an X.
The scale of the VT plots and their corresponding X-ray images is the same. The X-ray clusters can be easily identified in the high-density areas of the 
Voronoi plots. In the VT plots of XLSSC-d and XLSSC-f at least one more non-X-ray-detected overdensity 
can be seen, probably below the detection limit of the XXL Survey.}
\end{figure*}

Most optical counterparts of relatively bright X-ray sources ($f_{(0.5-2{\rm\; keV})}>1.0\times10^{-14}$ erg s$^{-1}$ cm$^{-2}$) in our three superclusters 
have spectroscopic redshifts. In more detail, in XLSSC-e five out of six sources have spectroscopy within a $10'$ radius of the centre of the
supercluster. Similarly, within the $25'$ of XLSSC-d and -f we find 38 out of 41, and 38 out of 51, respectively. On the contrary, 
only 4 out of the 13 sources below this flux have spectroscopy in XLSSC-e (we note that in XLSSC-e the flux limit is 
$f_{(0.5-2{\rm\; keV})}=4.5\times10^{-15}$ erg s$^{-1}$ cm$^{-2}$, see \S4.1).


The results of the current analysis
are based mainly on spectroscopic data, although the photometric redshifts of all sources were available. The optical counterparts of the 
sources with no available spectra are either too faint or totally absent and are therefore improbable  supercluster members. In fact,
studying their redshift probability distributions (PDZ), only one source in Field-3 is possibly at the redshift of the supercluster.  
Photometric redshifts were calculated for all our sources with suitable AGN and quasar templates (Fotopoulou et al. in prep).

In the case of the non-virialised superclusters the boundaries and
the geometry cannot be easily defined and the clusters that form the superclusters have a spread in redshift space.
Therefore, we initially based our selection of supercluster members on the condition $\delta z=|z_{spec}-z_{mean}|<2000(1+z_{cl})$ km/s, where
$z_{spec}$ is the galaxy redshift and $z_{mean}$ is the mean redshift of the supercluster members, which is a good 
approximation of more sophisticated cluster membership selection algorithms 
(e.g. Old et al. 2014 and references therein). 
Then, we extended the search for AGN to $1.5\times\delta z$ and $2\times\delta z$.

The expected spatial X-ray point-like density is calculated from the luminosity function of Hasinger, Miyaji \& Schmidt (2005). To
this end we first calculated the volume that is defined by the limits described in the previous paragraph. This is actually a cylinder
of volume $V$ given by $V=\pi R^{2}\times h$, where $R$ is the projected radius and $h$ is the height of the cylinder that corresponds
to the distance between the lower and upper redshift\footnote[6]{The height $h$ used in the above calculations is based on the selected $\delta z$ 
and is larger than $R$  so that the effect of galaxy peculiar velocities on the observed redshift distance between two sources is included.}. 
Then, we integrated the luminosity function within the luminosity range of interest
to calculate the expected number of sources per Mpc$^3$. Finally, by multiplying the two values we found the expected number of sources 
in the area of the superclusters. In all cases the expected number of X-ray point-like sources was less than one.

The results of the projected and the spatial X-ray overdensity analysis are summarised in Table 3.

\subsubsection{Optical galaxy spatial overdensity}

Any excess of X-ray point-like sources in the area of galaxy clusters can be due to the obvious abundance of galaxies with respect to the field
(see Koulouridis \& Plionis 2010). Therefore, to reach a meaningful interpretation of the X-ray point source overdensity analysis, and to reach a conclusion on
the enhancement or suppression of AGN,
we needed first to study the optical galaxy overdensity profile in the three fields. To this end, we used the photometric redshifts of the
CFHTLS-T0007 W1 field (Ilbert et al. 2006 and Coupon et al. 2009)
computed from three to five optical bands. The accuracy is 0.031 at $i<21.5$ and
reaches $\sigma_{\delta z/(1+z_{sp})} \sim 0.066$ at $22.5<i<23.5$. The fraction of
outliers increases from $\sim 2\%$ at $i<21.5$ to  $\sim$10 - 16\% at
$22.5<i<23.5$.

The relevant expression of the optical galaxy overdensity is similar to that of the 
X-ray overdensity in Eq. (1), i.e.
\begin{equation}
\;\;\;\;\;\;\;\;\;\;\;\;\;\;\;\;\;\;\;\;\;\;\;\delta_o=\frac{N_o}{N_{o, \rm exp}}-1\;,
\end{equation}
where $N_o$ is the number of optical sources found in the area and
$N_{o, \rm exp}$ the expected background number within the same area. 
For the calculation of the galaxy density, we considered the regions
previously defined for the X-ray analysis.
The expected galaxy density was calculated from a 2 deg$^2$ field within the
XMM-LSS area, free from clusters in the redshift range of the superclusters.

\subsection{Results}

From the 2D analysis of the $10'$ of XLSSC-e we expect $\sim$22 point-like sources above the lower flux limit, 
$f_{(0.5-2{\rm\; keV})}=4.5\times 10^{-15}$ erg s$^{-1}$ cm$^{-2}$, 
and we actually find 19. These numbers are consistent within the 1$\sigma$ confidence level (Gehrels 1986). 
In XLSSC-d and -e, $\sim$41 and $\sim$42 X-ray point-like sources are expected above
$f_{(0.5-2{\rm\; keV})}=1\times 10^{-14}$ erg s$^{-1}$ cm$^{-2}$, respectively. Indeed,
41 X-ray point-like sources are found in XLSSC-d. Nevertheless, in XLSSC-f we discover 51 and therefore a significant 
X-ray overdensity is found in the area, not consistent with the expected value at the 1$\sigma$ confidence level.

\begin{table}
\caption{2D and 3D analysis} 
\centering
\tabcolsep 5 pt
\def\arraystretch{1.2}
\begin{tabular}{lc|c|cccc} \hline
&&2D&\multicolumn{3}{c}{3D}\\
Name &  $R$ & N &$\pm\delta z$ & $\pm1.5\times\delta z$ & $\pm2\times\delta z$\\
  (1)    &  (2) &  (3)&  (4) & (5)& (6) \\
\hline
XLSSC-e & 10  & 19 (22)  &  2 ($<1$) & 2 ($<1$)& 2 ($<1$)\\   
XLSSC-d & 25 & 41 (41) &0 ($<1$)& 0 ($<1$)& 0 ($<1$)\\
XLSSC-f & 25 & 51 (42) &2 ($<1$) & 3 ($<1$)& 4 ($<1$)\\
\hline
\end{tabular}
\tablefoot{{\em (1)} Supercluster name; {\em (2)} projected search radius in arcmin; {\em (3)} number of detected X-ray point-like sources 
(in parentheses the expected number of sources calculated by the logN-logS); 
{\em (4)-(6)} number of AGN found within $1\times$,$1.5\times$, and $2\times\delta z$
of the supercluster redshift $z_{s}$, where $\delta z=\pm$2000(1+$z_{s}$) km/s (in parentheses the expected number of sources in the 
respective area calculated by the luminosity function).} 
\end{table}

Considering the available photometric and spectroscopic data (3D analysis within $\pm2\times\delta z$, see Table 3) 
we find in total six AGN with compatible 
spectroscopic redshifts in the area of the three superclusters 
(for the individual analysis see the next three paragraphs). 
The corresponding expected number of AGN calculated from the luminosity function (Hasinger, Miyaji \& Schmidt 2005) 
is $\sim$1.5 ($\sim$0.5 AGN per field) and therefore we calculate a total spatial AGN overdensity, $\delta_x=3$.  
As we have already pointed out, before reaching any definite conclusion we  also had to consider 
the high density of optical galaxies in the area of superclusters. Therefore, we also assessed the total 
spatial overdensity of optical galaxies in the three fields, $\delta_o=0.42$. We conclude that 
there is indeed a significantly higher overdensity of AGN with respect to the corresponding overdensity of 
optical galaxies at the 95\% confidence level (Gehrels 1986). This could indicate extra triggering of AGN caused by the environment.
We note that the above result is not affected by the selection of
a different redshift range ($\pm\delta z$ or $\pm1.5\times\delta z$, see Table 3);  although we detect fewer AGN, we 
also consider a much smaller volume.

Next, we proceed with the 3D analysis of each supercluster individually. We find
two AGN at the redshift of the XLSSC-e (Field-1) within $10'$
radius (Fig. 2, top). Both are low-luminosity AGN with $L_{(0.5-2{\rm\; keV})}\sim4\times10^{42}$erg s$^{-1}$.
The detailed overdensity of optical galaxies, divided into five annuli in each field, is plotted in Fig. 3.
It is apparent that in the area of XLSSC-e the bulk of the galaxies are concentrated in the central $4'$, 
while in the last annulus the galaxy density reaches the field level. The AGN overdensity in the 
whole field is significantly higher than the optical galaxy overdensity at the 90\% confidence level despite the 
small number statistics (Gehrels 1986). 
We also note that for the XLSSC-e, radio observations were obtained in XXL paper IX, but no large 
radio galaxies were found within the overdensities. They only associated eight radio sources with potential supercluster member galaxies; however, they are not  
associated with any of our X-ray point-like sources. 

Similarly, in XLSSC-f (Field-3) we find two to four spectroscopically confirmed AGN compatible with the supercluster redshift (depending on the 
redshift range, see \S4.1.2 and Table 3) and 
another possible member with compatible photometric redshift (Fig. 2, bottom). The optical galaxy overdensity profile (Fig. 3)
is almost flat over the whole field. This is probably due to cluster XLSSC 111, which is very massive with a large $r_{500,MT}$ 
radius and affects the full field. The X-ray overdensity is significantly higher than the optical overdensity at the 99\% 
confidence level. Similar to the two AGN found in XLSSC-e, the four AGN found in XLSSC-f are low-luminosity sources 
($L_{(0.5-2{\rm\; keV})}<7\times10^{42}$erg s$^{-1}$). The list of AGN can be found in Table 4. 

In contrast, no AGN is found within a $25'$ radius in XLSSC-d (Fig. 2, middle). In this field the large number of optical galaxies are 
located in annuli 2 to 4 (Fig. 3), as expected from the location of the three clusters, while in the first and last annuli the density 
reaches the expected field value. 
The overdensity of AGN in Field-2 is $\delta_x=-1$, but the null hypothesis that it is consistent with the optical galaxy overdensity
cannot be rejected at any statistically significant level.

\begin{table}
\caption{AGN in superclusters} 
\centering
\tabcolsep 3 pt
\def\arraystretch{1.2}
\begin{tabular}{lcccc} \hline
Name &  RA & DEC & $z$ & $L_{(0.5-2{\rm\; keV})}$ \\
  (1)    &  (2) &  (3)&  (4) & (5) \\
\hline
3XLSS J021046.2-060854 & 32.6928 & -6.1485  & 0.428 &  4.15$\times10^{42}$ \\   
3XLSS J021053.0-061809 & 32.7211 & -6.3026  & 0.423 &  3.66$\times10^{42}$ \\
3XLSS J021309.2-055142 & 33.2886 & -5.8618  & 0.298 &  6.03$\times10^{42}$ \\
3XLSS J021153.5-053810$\dagger$ & 32.9729 & -5.6363  & 0.288 &  3.58$\times10^{42}$ \\
3XLSS J021213.7-060408$\dagger$ & 33.0571 & -6.0690  & 0.283 &  6.21$\times10^{42}$ \\
3XLSS J021235.9-053210 & 33.1499 & -5.5364  & 0.299 &  6.70$\times10^{42}$ \\      
\hline
\end{tabular}
\tablefoot{{\em (1)} X-ray source name; {\em (2), (3)} field coordinates in the J2000 system; {\em (4)} redshift; {\em (5)} Soft X-ray luminosity in 
 units of erg s$^{-1}$.\\ $\dagger$ Included in the 1000 brightest XXL X-ray point source catalog (Fotopoulou et al. submitted, XXL paper VI) } 
\end{table}

We note that a low-luminosity source is actually detectable
in a smaller fraction of XLSSC-e compared to the other two superclusters because of their different redshift ($\sim$15\% smaller 
effective area for a source with $f_{(0.5-2{\rm\; keV})}\sim5\times 10^{-15}$ erg s$^{-1}$ cm$^{-2}$). In addition, the spectroscopic completeness in XLSSC-e
is less than 50\%, while in XLSSC-d and -f it is 93\% and 75\%, respectively. Therefore, there is some probability that we have missed  
supercluster members in XLSSC-e, although, as we have already discussed, the photometric redshift PDZs and the images of the optical counterparts
render this probability small. Only in XLSSC-f is the probability of one extra AGN  high, but in this supercluster the X-ray overdensity 
is already high without including the non-spectroscopic sources.  
 
The results imply some intrinsic differences between the superclusters. 
XLSSC-e includes five merging clusters in tight configuration, while 
the three clusters of XLSSC-f are not part 
of any further significant overdensity. Nevertheless, their total mass is  a factor of 2.5 
larger than that of the three clusters of XLSSC-d. On the other hand, although the three clusters of XLSSC-d 
form the less massive structure, they are part of a larger overdensity that includes  
at least another four clusters that we have not probed with the current observations.
Therefore, we cannot conclude in the current paper on the reasons that produce the observed differences because of the small number of superclusters 
studied. We will extend our study to the full XXL Survey in order to understand  these differences better and to quantify any trend regarding 
the AGN frequency in the environment of superclusters.

\begin{figure}
\resizebox{9cm}{11cm}{\includegraphics{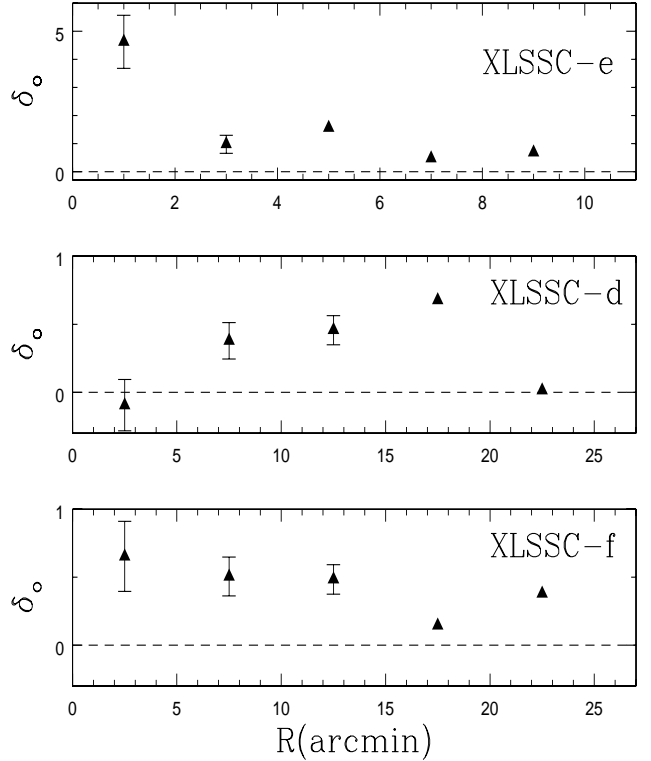}}
\caption{Spatial overdensity profile of optical galaxies within annuli of $2'$ (XLSSC-e) and $5'$ (XLSSC-d, XLSSC-f)
centred at the geometrical centre of the superclusters. For the calculation of the overdensity, CFHTLS
photometric redshifts of galaxies were used. The error-bars are $1\sigma$ Poissonian uncertainties and are shown
only for the first two or three annuli of each field. For the others the errors are smaller than the height of the
symbols.}
\end{figure}

\section{Conclusions}
 
In the first part of the current paper we presented the multifibre spectroscopic observations in three 0.55 deg$^2$ fields in the XXL-N survey with the 4.2m WHT.
Our targets were candidate member galaxies of clusters and optical counterparts
of X-ray point-like sources.
We obtained spectra for 455 galaxies, 56 of which are AGN. We determined the redshift of 25 clusters, 6 of which 
belong to the merging supercluster XLSSC-e at $z\sim0.43$, and confirmed  
2 more superclusters in looser configurations at $z\sim0.3$.

In the second part, we investigated the AGN frequency in
the environment of the superclusters. To this end, 
we identified all possible AGN supercluster members, which we define as sources
with $L_{(0.5-2{\rm\; keV})}>2.7\times10^{42}$ erg s$^{-1}$, and
compared their projected and spatial overdensity with the expected
overdensity of optical galaxies in the region. In more detail:

\begin{itemize}
 \item XLSSC-d: The supercluster presents no significant 2D overdensity of X-ray point-like sources 
and the total lack of AGN found by the 3D analysis is statistically consistent with the expected number of AGN within the area. 
\item XLSSC-f: In sharp contrast to XLSSC-d, a high projected overdensity of X-ray point-like sources 
was found by the 2D analysis. This result was confirmed by the 3D analysis, where the high number of spectroscopically confirmed AGN  
significantly exceeded the optical galaxy density expectations.
\item XLSSC-e: Similarly to XLSSC-f, we find a relatively high number of spectroscopically confirmed AGN that again
exceed the optical galaxy density expectations. The statistical significance of this result is not as high as for XLSSC-f and it is not supported 
by the 2D analysis. However, the probability that we have missed some AGN in this field is higher than in the other two fields.
\end{itemize}
Overall, the number of AGN in the area of the three superclusters significantly exceeds the field expectations
at the 95\% confidence level.

All six AGN found in the area of the superclusters have 
X-ray luminosities below $7\times10^{42}$ erg s$^{-1}$ and we can argue that they are low-luminosity sources. Similarly, a high number of low-luminosity 
AGN was reported in studies of AGN in rich clusters (e.g. Martini et
al. 2002, 2006; Davis et al. 2003), but not above the field expectations (e.g. Martini et
al. 2007; Haggard et al. 2010).  
In addition to our own data,  optical spectroscopy by the SDSS-BOSS project (Dawson et al. 2013) also exists for these sources. 
Except for 3XLSS J021235.9-053210, none of the AGN spectra presents broad permitted emission lines. A more thorough investigation of the AGN population 
in superclusters will be presented in a future paper.

The reason for the difference between the AGN frequency in the three superclusters cannot be completely 
understood by the present study because of the small sample.
To better understand the relation between AGN and the environment of superclusters, we will need to apply the same analysis 
to a larger number of massive formations. The 
wide area of the XXL Survey will soon give us the opportunity to realise this kind of study. 

\acknowledgements
We would like to thank the anonymous referee for the useful comments and suggestions.
XXL is an international project based around an XMM
Very Large Programme surveying two 25 $deg^2$
extragalactic fields at a depth of $5\times10^{-15}$ erg s$^{-1}$ cm$^{-2}$ in [0.5-2] keV at the 90\% completeness level (see XXL paper I).
The XXL website is
http://irfu.cea.fr/xxl. Multiband information and spectroscopic follow-up of the
X-ray sources are obtained through a number of survey programmes, summarised at http://xxlmultiwave.pbworks.com/.
The authors acknowledge L. Chiappetti for producing a format and catalog compliance report. 
EK acknowledges fellowship funding provided by the Greek General Secretariat of Research
and Technology in the framework of the programme Support of Postdoctoral
Researchers, PE-1145. EK acknowledges the Centre National d’Etudes Spatiales 
(CNES) and CNRS for support of post-doctoral research. YJ acknowledges support by FONDECYT grant N. 3130476.
O.M. is grateful for the financial support provided by the NEWFELPRO fellowship project in Croatia.
FP acknowledges support from the BMBF/DLR grant 50 OR 1117,
the DFG grant RE 1462-6 and the DFG Transregio Programme TR33.
This work is based on observations obtained with MegaPrime/MegaCam, a joint project of CFHT
and CEA/IRFU, at the Canada-France-Hawaii Telescope (CFHT) which is operated by
the National Research Council (NRC) of Canada, the Institut National des Sciences
de l'Univers of the Centre National de la Recherche Scientifique (CNRS) of
France, and the University of Hawaii. This work is based in part on data
products produced at Terapix available at the Canadian Astronomy Data Centre as
part of the Canada-France-Hawaii Telescope Legacy Survey, a collaborative
project of NRC and CNRS.

\end{document}